\documentclass[twocolumn,superscriptaddress]{revtex4}

\usepackage{xcolor}
\usepackage{soul}
\usepackage{graphicx}
\usepackage{dcolumn}
\usepackage{bm}
\usepackage{color}
\usepackage{amsmath,amssymb}
\usepackage{physics}
\usepackage{ulem}

\usepackage{here}

\def\gesim{\ \raise.3ex\hbox{$>$}\kern-0.8em\lower.7ex\hbox{$\sim$}\ }

\newcommand{\nn}{\nonumber}
\newcommand{\sei}[2]{\hat{c}^{#1}_{#2}}

\newcommand{\tsei}[2]{\hat{F}^{#1}_{#2}}
\newcommand{\bk}{\bm{k}}

\begin{document}
\title{Condensation of Cooper Triples}
\author{Sora Akagami}
\affiliation{Department of Mathematics and Physics, Kochi University, Kochi 780-8520, Japan}
\author{Hiroyuki Tajima}
\affiliation{Department of Mathematics and Physics, Kochi University, Kochi 780-8520, Japan}
\affiliation{Department of Physics, Graduate School of Science, The University of Tokyo, Tokyo 113-0033, Japan}
\author{Kei Iida}
\affiliation{Department of Mathematics and Physics, Kochi University, Kochi 780-8520, Japan}

\date{\today}
\begin{abstract}

{The condensation of Cooper pairs, originating from the Fermi-surface instability due to a weakly attractive interaction between two fermions, opened a new frontier for exploring many-body physics in interdisciplinary contexts.
In this work, we discuss the possible condensation of Cooper triples, which are three-body counterparts of Cooper pairs for three-component fermions with a three-body attraction.
Although each composite trimer-like state obeys the Fermi-Dirac statistics, its aggregate can form a condensate at zero center-of-mass momentum in the presence of the internal degrees of freedom associated with the relative momenta of constituent particles of momenta close to the Fermi surface.
Such condensation can be regarded as bosonization in {\it infinite}-component fermions.
We propose a variational wave function for the condensate of Cooper triples in analogy with the Bardeen-Cooper-Schrieffer ground state and obtain the ground-state energy.
}
\end{abstract}
\maketitle
\noindent
{{\it Introduction}---}
A quantum many-body problem is related to various systems that are encountered not only in condensed matter physics but also in nuclear and particle physics.
 On top of interactions between constituents, the degrees of freedom such as spin, isospin, and flavor play an important role in characterizing a variety of strongly correlated systems.
One of the most striking examples is conventional superconductivity, which is triggered by the formation of a Cooper pair, i.e., a pairing state of two electrons with spin up and down in the presence of the Fermi sphere.
The Bardeen-Cooper-Schrieffer (BCS) theory~\cite{1}, which successfully explains the microscopic mechanism of superconducting phase transition in terms of condensation of Cooper pairs,
has developed a fundamental basis for the description of superfluid states in cold atoms~\cite{Bloch,Giorgini} as well as dense matter~\cite{Dean,Alford}.


 Most of quantum many-body effects in spin-$1/2$ fermions  (e.g., electrons, neutrons) have been studied in terms of two-body interactions as in the BCS theory~\cite{1}. 
Three or even more body interactions between particles with internal multiple degrees of freedom, on the other hand, have attracted a lot of attention~\cite{2}.
In a three-component fermionic system, the formation of Cooper triples, which correspond to a three-body version of Cooper pairs, has been studied in the presence of two-body and three-body interactions~\cite{9,TajimaTriple,Kirk,10}.
If Cooper triples actually occur, however, what kind of state the system
would end up with is not known even qualitatively and hence to be addressed
here.
It is also under investigation how medium corrections affect properties of Efimov trimers~\cite{MacNeill,Nishida,Nygaard,Tajima1,23,Pierce,Sanayei1,Sanayei2}.
In addition, an in-medium cluster state consisting of more than three particles such as a Cooper quartet has also been pointed out~\cite{11,12}. Similarly, the alpha-particle condensation in nuclear matter has been investigated by generalizing the BCS ground state~\cite{13,Senkov,Baran,Baran2}.

 Although more than two-body interactions are generally weak and are often treated as small perturbations in binding energies dominated by two-body interactions~\cite{2}, such multi-body interactions are inherent to composite particles~\cite{3} and sometimes are needed to clarify the properties of in-medium cluster states.
 From this perspective, ultracold atoms provide a valuable opportunity to manipulate multi-body interactions in a systematic manner~\cite{4}.
 In particular, by reducing the two-body interaction to zero, one can realize a unique system that is governed by the three-body interaction hidden behind the two-body interaction~\cite{5,6,7,8,TajimaTriple,Kirk,Guijaro,Drut2018,Pricoupenko2,Valiente1,Pricoupenko,Valiente2,Valiente3,Colussi}.
  Effects of the three-body interaction have been studied extensively in theories of, e.g., pair superfluidity of bosons~\cite{14,15,16,17}, three-component Fermi gases~\cite{18,19,20}, quantum bound states~\cite{21}, and Efimov states~\cite{22,23}.
\begin{figure}[t]
\begin{center}
\includegraphics[width=7.0cm]{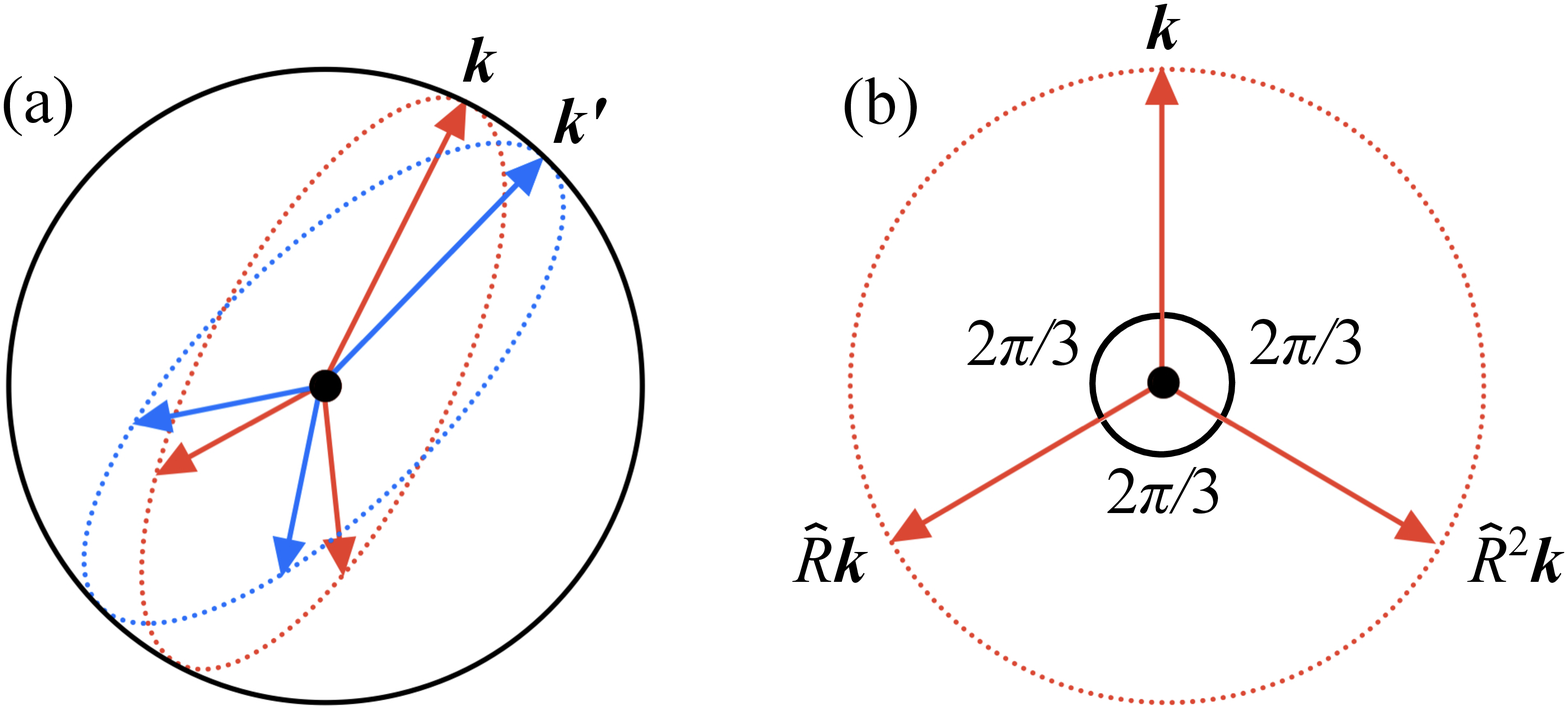}
\caption{(a) Momentum-space configurations of three fermions of different species on a spherical surface that have zero center-of-mass momentum. The momenta $\bk$ and $\bk'$ of one of the three fermions can take on various directions.
(b) The corresponding in-plane configuration of the three fermions of momenta
$\bk$, $\hat{R}\bk$, and $\hat{R}^{2}\bk$.
The angle between any pair of momenta is $2\pi/3$.
}
\label{fig:configuration}
\end{center}
\end{figure}

In this work, we theoretically predict that Cooper triples can condense in a three-component Fermi system.  Although each Cooper triple obeys the Fermi-Dirac statistics, a macroscopic number of triple states can occupy an effective single state of zero center-of-mass momentum in the presence of various sets of three particles having different species and different momenta on a spherical surface close to the Fermi surface as shown in Fig.~\ref{fig:configuration}.
This exotic condensation can be regarded as bosonization in {\it infinite}-component fermions associated with relative momenta.
For an attractive three-body force which has extensively been discussed in cold atom physics~\cite{2},
we propose a variational wave function for the Cooper triple condensation in analogy with the BCS ground state.
We obtain the ground-state energy in a three-component fermionic system with a three-body interspecies attraction to show that the Cooper triple state is energetically more favorable than the normal state.
In what follows, we take the system volume to be unity and units in which $\hbar=k_{\rm B}=1$.
\\

\noindent
{{\it Model}---}
We consider non-relativistic three-component (r,g,b) fermions of equal mass $m$ with the three-body attractive interaction.
Various systems with three-body interaction have been discussed in ultracold atoms~\cite{5,6,7,8,Guijaro,Pricoupenko2,Drut2018,Valiente1,Pricoupenko,Valiente2,Valiente3,Colussi}.
Moreover, we later propose a way to realize an attractive three-body interaction induced by a background medium gas~\cite{Frieman,Tajima2021,DeSalvo2019,Chui2004,Belemuk2007} in multi-component ultracold atomic gases (e.g. $^6$Li, $^{173}$Yb).
In this work, we assume that the two-body interaction is vanishing for simplicity, while the weakly attractive three-body interaction ignored in the previous work~\cite{9} is present.
It is noteworthy that even in the presence of the two-body attractive interaction alone among the three components,
the Cooper triple state is predicted to be dominant over the Cooper pairing state at least above a threshold strength~\cite{9}.
We also consider the system in which the chemical potentials are the same, i.e., $\mu_{\rm r}=\mu_{\rm g}=\mu_{\rm b}\equiv\mu$.
\par
At sufficiently low temperature and weak coupling, the Fermi degeneracy is expected and therefore the available momentum space of fermions undergoing zero
center-of-mass-momentum three-body scattering is restricted to around the Fermi surface.
In such a case, the system can be described by the following effective Hamiltonian in the momentum space,
\begin{align}
\label{Hamiltonian}
\hat{H}_{\rm eff}=\sum_{\gamma}\sum_{\bk}\xi_{\bk}\hat{n}_{\bk,\gamma}+\sum_{\bk,\hat{R}}\sum_{\bk',\hat{R'}}U_{\bk,\bk'}\tsei{\dag}{\bk',\hat{R'}}\tsei{}{\bk,\hat{R}}.
\end{align}
Here $\xi_{\bk}=|\bk|^2/(2m)-\mu$ is the kinetic energy of a fermion with momentum $\bk$ measured from $\mu$, and
$\hat{n}_{\bm{k},\gamma}=c_{\bm{k},\gamma}^\dag c_{\bm{k},\gamma}$ is the single-particle number operator with the annihilation (creation) operator $c_{\bm{k},\gamma}^{(\dag)}$ for fermions of momentum $\bk$ and component $\gamma$.
The second term in Eq.\ (\ref{Hamiltonian}) denotes the three-body interaction with a contact-type coupling constant $U$, taken to be negative here. $\tsei{(\dag)}{\bk,\hat{R}}$ is the trimer annihilation (creation) operator defined in terms of $c_{\bk,\gamma}$ as
\begin{align}
\label{eq:FF}
\tsei{\dag}{\bk,\hat{R}}=\sei{\dag}{\bk,{\rm r}}\sei{\dag}{\hat{R}\bk,{\rm g}}\sei{\dag}{\hat{R}^{2}\bk,{\rm b}},\quad
\tsei{}{\bk,\hat{R}}=\sei{}{\hat{R}^{2}\bk,{\rm b}}\sei{}{\hat{R}\bk,{\rm g}}\sei{}{\bk,{\rm r}},
\end{align}
where $\hat{R}$ is the operator that represents a $2\pi/3$ rotation in a given momentum plane and hence ensures zero center-of-mass momentum ($\bk+\hat{R}\bk+\hat{R}^{2}\bk=\bm{0}$).
These operators satisfy the anti-commutation relations
\begin{align}
\label{eq:anti}
\{\tsei{}{\bk,\hat{R}},\tsei{}{\bk',\hat{R'}}\}&=\{\tsei{\dag}{\bk,\hat{R}},\tsei{\dag}{\bk',\hat{R'}}\}=\hat{0},\\
\{\tsei{}{\bk,\hat{R}},\tsei{\dag}{\bk',\hat{R'}}\}&=\delta_{\bk,\bk'}\delta_{\hat{R}\bk,\hat{R'}\bk'}\delta_{\hat{R}^{2}\bk,\hat{R'}^{2}\bk'}\nn\\
&\times\left[ (\hat{1}-\hat{n}_{\bk,r})(\hat{1}-\hat{n}_{\hat{R}\bk,g})(\hat{1}-\hat{n}_{\hat{R}^{2}\bk,b})\right.\cr
&+\left.\hat{n}_{\bk,r}\hat{n}_{\hat{R}\bk,g}\hat{n}_{\hat{R}^{2}\bk,b} \right].
\label{eq:anti2}
\end{align}
Physically, 
the two terms in the right side of Eq.\ (\ref{eq:anti2}) represent a deviation from the usual fermionic anti-commutation relation due to a composite nature of three holes and three particles; for the normal state at zero temperature, such a deviation vanishes.\\

\noindent
{{\it Possible condensation of Cooper triples}---}
Let us proceed to ask how condensation of Cooper triples 
can occur by considering
a three-fermion configuration among different components as shown in Fig.~\ref{fig:configuration}.
For such three fermions near the Fermi surface, the absolute values of the respective momenta $|\bk|$, $|\hat{R}\bk|$, and $|\hat{R}^{2}\bk|$ are around the Fermi momentum $k_{\rm{F}}$, while the center-of-mass momentum remains zero.
In momentum space, as depicted in this figure, $\bk$, $\hat{R}\bk$, and $\hat{R}^{2}\bk$ are located on the same plane in such a way that the angles between two of them are $2\pi/3$.  When $\bk$ and the plane are fixed, therefore, the  other two, $\hat{R}\bk$ and $\hat{R}^{2}\bk$, are automatically determined.
\par
From the anti-commutation relation associated with $c_{\bm{k},\gamma}$, one finds
\begin{align}
    F_{\bm{k},\hat{R}}^\dag F_{\bm{k}',\hat{R}'}^\dag\ket{0}\neq 0
    \quad (\bm{k}\neq\bm{k}', \hat{R}\bm{k}\neq\hat{R'}\bm{k}', \hat{R}^2\bm{k}\neq\hat{R'}^2\bm{k}'),
\end{align}
where $\ket{0}$ is the normalized vacuum state. This indicates that two trimers with zero center-of-mass momentum can coexist unless the two momenta of a given component happen to be the same.
Since $\bk$ can be taken in countless ways, one can find a countless number of configurations of the three fermions with zero center-of-mass momentum in the presence of the Fermi surface.
It looks as if bosonization occurred in {\it infinite}-component fermions~\cite{24}.
Indeed, it is similar to the case of a three-dimensional gas of SU($N$) fermions in which the internal spin degrees of freedom of composite particles arise from the spin of constituent particles.
Therefore, the Cooper triples can condense by occupying the zero-momentum state macroscopically regardless of their Fermi-Dirac statistics.\\
\\

\noindent
{{\it Generalized Cooper problem}---}
 Before addressing whether Cooper triples actually condense, we have to confirm that a single Cooper triple appears as a bound state in the presence of the Fermi sea.  To this end,
we generalize the Cooper problem \cite{CooperProblem} to the present three-component case.
\par
First, we assume  that a trial wave function of a  single three-body state above the Fermi surface is given by 
\begin{align}
\label{WavFuncCooper}
    |\psi'\rangle=\sum_{|\bm{p}|\geq k_{\rm F}}\Phi_{\bm{p}}F_{\bm{p},\hat{R}}^\dag|\psi_{\rm FS}\rangle,
\end{align}
where
$|\psi_{\rm FS}\rangle$ denotes the ground state of noninteracting fermions  that corresponds to the Fermi sphere of radius $k_{\rm F}$, and $\hat{R}$ is fixed.
Variation of the expectation value $\langle\psi'|\hat{H}_{\rm eff}|\psi'\rangle$ with respect to $\Phi_{\bm{p}}^*$ under $\langle\psi'|\psi'\rangle=1$ leads to a Schr\"{o}dinger-like equation, 
\begin{align}
\label{eq:SCHR}
    \left(\frac{3p^2}{2m}-E-3E_{\rm F}\right)\Phi_{\bm{p}}=-\sum_{|\bm{p'}|\geq k_{\rm F}}U_{\bm{p'},\bm{p}}\Phi_{\bm{p'}},
\end{align}
where $E_{\rm F}=\frac{k_{\rm F}^2}{2m}$ is the Fermi energy.
\par
As in the case of the usual Cooper problem, the existence of the $E<0$ solution to Eq.\ (\ref{eq:SCHR}) indicates that the three-body state is bound due to the Fermi surface effect. In the present case, we assume
\begin{align}
\label{eq:U}
U_{\bm{k},\bm{k}'}=-U_0\theta(\Lambda-|\xi_{\bm{k}}|) \theta(\Lambda-|\xi_{\bm{k}'}|),
\end{align}
where $U_0$ is the positive constant. 
In the presence of the Fermi surface common to the three components,
we introduce an energy cutoff $\Lambda$ such that the interaction works only for three fermions of momenta close to the Fermi surface.
We note that $\Lambda$ corresponds to the Debye frequency in a conventional BCS superconductor with phonon-mediated interaction~\cite{1}.
In the case of the fermion-mediated three-body interaction, as will be explained later,
$\Lambda$ is associated with the Fermi energy of medium fermions~\cite{56}.
By incorporating Eq.\ (\ref{eq:U}) into Eq.\ (\ref{eq:SCHR}), one can obtain
the equation for the three-body energy $E$ as 
\begin{align}
    1=U_0\sum_{|\bm{p}|\geq k_{\rm F}}\frac{\theta(\Lambda-|\xi_{\bm{p}}|)}{\frac{3p^2}{2m}-E-3E_{\rm F}}.
\end{align}
In the weak-coupling limit, one arrives at $E=-3\Lambda\exp\left(-\frac{3}{\rho(0)U_0}\right)<0$, where $\rho(0)$ is the density of state at the Fermi level.
In this way, one can conclude that the unperturbed ground state is unstable with respect to the formation of a Cooper triple. 

We emphasize that this instability originates from the Fermi surface effect since there are no bound states in vacuum in the case of an infinitesimal small three-body coupling. 
Indeed, by following a line of argument of Ref.\ \cite{10}, 
the threshold for the occurrence of a three-body bound state is found to be nonzero.
Although these in-vacuum quantities, together with the in-medium ones, may be renormalized by physical quantities such as the three-body scattering amplitude if one goes beyond the weak-coupling analysis, in this work we specifically focus on the weak-coupling limit.

\begin{figure}[t]
\begin{center}
\includegraphics[width=6cm]{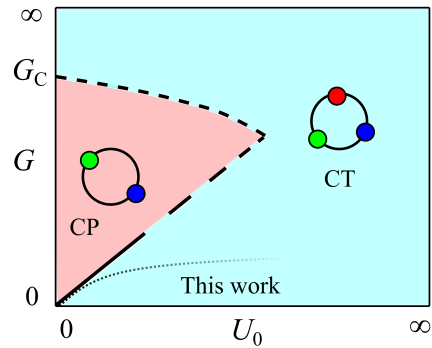}
\caption{Schematic phase diagram in the plane of the two-body coupling constant $G$ and three-body coupling constant $U_0$ (in arbitrary units).
The solid line shows the threshold for the transition between a Cooper pair (CP) and a Cooper triple (CT) state as obtained within the generalized Cooper problem in the weak-coupling limit. $G_{\rm C}$ is the critical two-body coupling constant for the transition between CP and CT in the absence of $U_0$~\cite{9}.
We also plot an eye-guide (dashed line) for the upper bound of CP starting from $(G,U_0)=(G_{\rm C},0)$.
In this work, we consider the weak-coupling region where the Cooper triple formation is dominant as indicated below the dotted line.}
\label{fig:PhaseDiagram}
\end{center}
\end{figure}

Let us now show where the situation of interest here is located in the space of the two-body and three-body coupling constants.
Figure~\ref{fig:PhaseDiagram} exhibits a 
schematic phase diagram based on weak-coupling analyses of the Cooper problem in the presence of the two-body attractive interaction also.
In such a case, a Cooper pair can be formed in the unperturbed ground state.
The binding energy of such a Cooper pair is given by $E_2=-2\Lambda\exp\left(-\frac{2}{\rho(0)G}\right)$,
where {$G\geq 0$ is the attractive} two-body coupling constant (for simplicity, we take the same cutoff $\Lambda$ as the three-body case). We can thus conclude from comparison between the in-medium two-body and three-body binding energies that
 the transition between a Cooper pair and a Cooper triple state occurs at $G=\frac{2}{3}U_0$ in the weak-coupling limit.
In this work, we focus on the weak-coupling region where the Cooper triple formation is dominant due to $U_0\gg G$ (indicated by the region below the dotted line in Fig.~\ref{fig:PhaseDiagram}).
We note that the transition from the Cooper pair to the Cooper triple state in the absence of the three-body attraction requires a finite strength of the two-body attraction~\cite{9,TajimaTriple}, which corresponds to the critical coupling $G_{\rm C}$ in Fig.~\ref{fig:PhaseDiagram}.
The upper bound of the Cooper pair state at finite $U_0$ (eye-guide denoted by the dashed line), can be complicated by the coexistence or competition of two-body and three-body correlations and hence is out of scope in this paper.\\

\noindent
{\it Variational wave function for the Cooper triple condensation}---
Since the generalized Cooper problem as discussed above suggests the new ground state involving many Cooper triples,
we propose a gauge fixed variational wave function for the Cooper triples, in analogy with the BCS ground state, as
\begin{align}
\label{WaveFunction}
\ket{\psi_T}=\prod_{\bk}(u_{\bk}+v_{\bk}\tsei{\dag}{\bk,\hat{R}_0})\ket{0},
\end{align}
where $u_{\bk}$ and $v_{\bk}$ are complex variational parameters as functions of $\bk$ alone.
We note that the state (\ref{WaveFunction}) breaks the $U(1)$ gauge symmetry via the superposition of Cooper triples. This fact is in sharp contrast to the case of bosonized SU($N$) fermions~\cite{24} where the phase coherence is absent since the lowest energy level is simply occupied by each hyperfine state.
Here, we set $\hat{R}_0$ in such a way that if the momentum $\bk$ of component r is in the direction of ${\bf e}_1=\left(\begin{array}c 0\\0\\1\end{array}\right)$, the momentum $\hat{R}_0\bk$ of component g and the momentum $\hat{R}_0^2\bk$ of component b are in the
direction of ${\bf e}_2=\left(\begin{array}c \sqrt3/2\\0\\-1/2\end{array}\right)$ and ${\bf e}_3=\left(\begin{array}c -\sqrt3/2\\0\\-1/2\end{array}\right)$, respectively; otherwise, $\bk$, $\hat{R}_0\bk$, and $\hat{R}_0^2\bk$ are in the direction of $V{\bf e}_1$, $V{\bf e}_2$, and $V{\bf e}_3$, respectively,
with an appropriate rotation matrix $V$.
This choice of $\hat{R}_0$, leading to a specific orientation in momentum space, ensures that the state (\ref{WaveFunction}) can have any momentum of each component picked up only once from the vacuum.  
Note that one can consider another $\hat{R}_0$ by setting ${\bf e}_2=\left(\begin{array}c \sqrt3\cos\theta/2 \\ \sqrt3\sin\theta/2\\-1/2\end{array}\right)$ and ${\bf e}_3=\left(\begin{array}c -\sqrt3\cos\theta/2\\ -\sqrt3\sin\theta/2\\-1/2\end{array}\right)$ with $\theta\neq0$, i.e., by rotating the momentum plane on which a Cooper triple with $\bk$ in the direction of ${\bf e}_1$ resides by $\theta$ with respect to ${\bf e}_1$.  The resultant state is degenerate with the original state. Such degeneracy is similar to the case of different gauge orientations, which allows us to regard the variational parameters as independent of the choice of $\hat{R}_0$ as well as the gauge.  Note also that in the BCS case $\bk$ just corresponds to the relative momentum of a Cooper pair, whereas in the state (\ref{WaveFunction}) not only $\bk$ but also $\hat{R}_0$ characterizes the relative momenta of two fermions in a Cooper triple: $(1-\hat{R}_0)\bk$, $(\hat{R}_0-\hat{R}_0^2)\bk$, and $(\hat{R}_0^2-1)\bk$.
\par
One may consider a more sophisticated variational wave function that involves the superposition of the states given by Eq.\  (\ref{WaveFunction}) with various $R_0$'s.
Since the variational parameter space is enlarged by considering such superposition, the ground-state energy in this wave function would be lower than that of the state (\ref{WaveFunction}).
For our purpose, however, it is enough to show that the ground-state energy is lowered compared to the normal state by the Cooper triple formation considered in Eq.~(\ref{WaveFunction}).
Since we consider the infinitesimally weak three-body attraction in Eq.~(\ref{Hamiltonian}), moreover, 
a zero center-of-mass momentum configuration of Cooper triples 
is restricted to $(\bm{k},{\rm r})$, $(\hat{R}_0\bm{k},{\rm g})$, and $(\hat{R}_0^2\bm{k},{\rm b})$ on the Fermi surface. For finite coupling, it is possible to construct a variational wave function composed of the creation operator $F^\dagger$ of a different kind of three-fermion configuration with zero total momentum, but we have difficulty in finding a variational wave function that can describe not only a condensed state but also the Fermi sphere common to the three components. This is because once for $F^\dagger$ one chooses a specific triangle that has a center of mass at the origin but has different lengths from the origin among the three components in momentum space, then one would fail to describe the Fermi sphere and hence a second order transition to a condensed state.

\par
Under $\bra{\psi_T}\ket{\psi_T}=1$, the normalization condition of $u_{\bk}$ and $v_{\bk}$ reads
$|u_{\bk}|^2+|v_{\bk}|^2=1$,
where $|u_{\bk}|^2$ and $|v_{\bk}|^2$ physically represent the unoccupied and occupied probabilities of a Cooper triple with $(\bk,{\rm r})$, $(\hat{R}_0\bk,{\rm g})$, and $(\hat{R}_0^{2}\bk,{\rm b})$,
respectively.
{Then, using Eqs.\ (\ref{eq:anti}) and (\ref{eq:anti2}), one can evaluate the ground-state energy $E_0=\bra{\psi_T}\hat{H}_{\rm eff}\ket{\psi_T}$ as}
\begin{align}
\label{GroundStateEnergy}
E_0=\sum_{\bk}3\xi_{\bk}|v_{\bk}|^2
+\sum_{\bk}\sum_{\bk'}U_{\bk,\bk'}v^{\ast}_{\bk}v_{\bk'}u^{\ast}_{\bk'}u_{\bk}.
\end{align}
Minimization of $E_0$ with respect to the variational parameters leads to
{$u_{\bk}=\frac{1}{\sqrt{2}}\left(1+\frac{\xi_{\bk}}{\sqrt{\xi^{2}_{\bk}+\Delta^{2}_{\bk}}}\right)^{1/2},
v_{\bk}=\frac{1}{\sqrt{2}}\left(1-\frac{\xi_{\bk}}{\sqrt{\xi^{2}_{\bk}+\Delta^{2}_{\bk}}}\right)^{1/2}$},
where $\Delta_{\bk}\equiv-\frac{2}{3}\sum_{\bk'}U_{\bk,\bk'}u_{\bk'} v_{\bk'}$ is the order parameter characterizing the Cooper triple condensation. This is because $\Delta_{\bk}$ can be written as
\begin{align}
\label{eq:op}
  \Delta_{\bk}
  =-\frac{2}{3}\sum_{\bk'}U_{\bk,\bk'} \langle c_{\bm{k}',{\rm r}}^\dag c_{\hat{R_0}\bm{k}',{\rm g}}^\dag c_{\hat{R_0}^{2}\bm{k}',{\rm b}}^\dag\rangle.
\end{align}
$\Delta_{\bm{k}}$ is the expectation value of the fermionic operator,
which is in sharp contrast to the BCS superconducting gap~\cite{1}.

Before calculating the order parameter, we consider the non-interacting case by setting $\Delta_{\bm{k}}\rightarrow 0$.
In such a case, we obtain $u_{\bm{k}}=\theta(k_{\rm F}-|\bm{k}|)$ and $v_{\bm{k}}=\theta(|\bm{k}|-k_{\rm F})$ 
, where $\theta(x)$ is the step function.
This result indicates that Eq.~(\ref{WaveFunction}) reproduces the wave function corresponding to the filled Fermi sphere, which is given by
\begin{align}
\label{eq:FS}
\ket{\psi_{\rm FS}}=\prod_{|\bk|\leq k_{\rm F}}\tsei{\dag}{\bk,\hat{R}_0}\ket{0}\equiv \prod_{\gamma}\prod_{|\bk|\leq k_{\rm F}}c_{\bm{k},\gamma}^\dag \ket{0}.
\end{align}


Let us move on to the weakly interacting case.
{From Eq.~(\ref{GroundStateEnergy}), we obtain}
\begin{align}
\label{eq:GSE}
E_0&=\frac{3}{2}\sum_{\bk}\xi_{\bk}\left(1-\frac{\xi_{\bk}}{\sqrt{\xi^2_{\bk}+\Delta^2_{\bk}}}\right)\nn\\
&+\frac{1}{4}\sum_{\bk}\sum_{\bk'}U_{\bk,\bk'}\frac{\Delta_{\bk}}{\sqrt{\xi^2_{\bk}+\Delta^2_{\bk}}}\frac{\Delta_{\bk'}}{\sqrt{\xi^2_{\bk'}+\Delta^2_{\bk'}}}.
\end{align}
Substituting Eq.~(\ref{eq:U}) to Eq.~(\ref{eq:op}), we obtain
  $\Delta_{\bm{k}}=\Delta\theta(\Lambda-|\xi_{\bm{k}}|)$,
where $\Delta$ is the amplitude of the order parameter.
In the weak coupling limit ($\Delta\ll\Lambda$), as in the BCS theory, $\Delta$ can be analytically obtained as
\begin{align}
\label{eq:DeltaOrderParameter}
\Delta\approx2\Lambda\exp\left(-\frac{3}{2\rho(0)U_0}\right),
\end{align}
where
$\rho(\omega)
=\frac{1}{(2\pi)^2}(2m)^{3/2}(\omega+\mu)^{1/2}$
is the density of state as a function of the single-particle energy $\omega$ in an ideal Fermi gas. 
By combining Eqs.\ (\ref{eq:U}) and (\ref{eq:GSE}), we can express the ground-state energy as
\begin{align}
\label{eq:GSEcutoff}
E_0\simeq E_0^{\rm FS}+\frac{3}{2}\sum_{|\xi_{\bk}|\leq \Lambda}|\xi_{\bk}|\left(1-\frac{|\xi_{\bk}|}{\sqrt{\xi^2_{\bk}+\Delta^2}}\right)
-\frac{9}{4}\frac{\Delta^2}{U_0},
\end{align}
where we split the summation in Eq.~(\ref{eq:GSE}) as $\sum_{\bm{k}}=\sum_{|\xi_{\bk}|> \Lambda}+\sum_{|\xi_{\bk}|\leq \Lambda}$.
While the former gives a large part of the Fermi-gas energy $E_0^{\rm FS}$ since $U_{\bm{k},\bm{k}'}=0$ there,
the latter is responsible for the interaction effect near the Fermi surface.
Under the assumption of $\Delta\ll \Lambda$,
the difference in the ground-state energy between the Cooper triple and Fermi degenerate states is given by
\begin{align}
\label{eq:ConEne}
E_0-E_{0}^{\rm FS}
&\approx -\frac{3}{4}\rho(0)\Delta^2 \cr
&\approx
-3\rho(0)\Lambda^2\exp\left(-\frac{3}
{\rho(0)U_0}\right) < 0.
\end{align}
Whenever $U_0$ is nonzero, therefore, the ground-state energy of the Cooper triple state is lower than that of the normal Fermi gas.
In addition, Eq.\ (\ref{eq:ConEne}) can be regarded as the condensation energy of the Cooper triple state as in the BCS ground state with Cooper pairs~\cite{1}.\\

\noindent
{\it Mediated two- and three-body interactions}---
\label{sec:6}
We discuss the possibility of realizing a mediated three-body attractive interaction in multi-component quantum gases.
We consider three-component fermions immersed in a medium atomic gas, which can be realized in four-component mixtures.
The candidates could be a three-component mixture of $^6$Li or $^{173}$Yb immersed in a background majority gas that consists of another hyperfine state or atomic species.
Although either bosons or fermions will do for medium atoms,
we consider a fermionic medium for the sake of the stability of the system. 
Since Cooper triples can occur even if the three-body attraction is 
infinitesimally weak, we determine the mediated two- and three-body interactions $U_{\gamma\gamma'}^{\rm eff}$ and $V_{\rm rgb}^{\rm eff}$ within the weak-coupling perturbation with respect to the fermion-medium coupling $U_{\gamma{\rm 4}}$~\cite{Tajima2021}.
In the low-energy and low-momentum limit at zero temperature,  the effective two-body interaction $U_{\gamma\gamma'}^{\rm eff}$ reads
\begin{align}
\label{eqV2}
    U_{\gamma\gamma'}^{\rm eff}
    &=U_{\gamma\gamma'} -U_{\gamma{\rm 4}}U_{\gamma'{\rm 4}}\frac{m_{\rm 4}k_{\rm F,4}}{2\pi^2},
\end{align}
where $U_{\gamma\gamma'}$ is the direct interaction between $\gamma$ and $\gamma'$ components. 
In Eq.~(\ref{eqV2}), $k_{\rm F,4}$ and $m_4$ are the Fermi momentum and mass of medium atoms, respectively.
As in the case of polaron problems, the backaction from the gas of three-component fermions to the medium can be neglected in the presence of the large population imbalance~\cite{Tajima2021}.
The second term in the right side of Eq.~(\ref{eqV2}) is consistent with the recent experiment~\cite{DeSalvo2019}.
$U_{\gamma\gamma'}^{\rm eff}$ becomes zero when the direct repulsive and the induced attractive interactions cancel each other.
Note, however, that the vanishing two-body interaction is not the necessary condition for the emergence of Cooper triples~\cite{9}.  
We also remark that we ignore the higher partial-wave components, which are negligible at low temperature.
\par
Next, we consider the mediated three-body interaction $V_{\rm rgb}^{\rm eff}$ up to leading order in $U_{\gamma{\rm 4}}$.
At zero temperature, we obtain
\begin{align}
\label{eqV3}
    V_{\rm rgb}^{\rm eff}
    &=U_{{\rm r}{\rm 4}}U_{{\rm g}{\rm 4}}U_{{\rm b}{\rm 4}}\frac{m_{\rm 4}^2}{2\pi^2 k_{\rm F,4}}. 
\end{align}
The sign of $V_{\rm rgb}^{\rm eff}$ depends on $U_{\gamma{\rm 4}}$ in such a way that $V_{\rm rgb}^{\rm eff}$ becomes negative when each or only one of the three $U_{\gamma{\rm 4}}$'s is negative.
For instance, in the case of $U_{{\rm r}{\rm 4}}>0$, $U_{{\rm g}{\rm 4}}>0$, and $U_{{\rm b}{\rm 4}}<0$,
we do obtain the attractive three-body interaction, i.e., 
$V_{\rm rgb}^{\rm eff}<0$.
We note that Eq.~(\ref{eqV3}) is consistent with the path integral result for a Bose-Fermi mixture~\cite{Chui2004,Belemuk2007}.
In this way, one can realize the mediated three-body attraction by tuning the bare two-body interactions in an appropriate way.
\par
As shown in Ref.~\cite{56}, $\Lambda$ in the fermion-mediated interaction is of the order of the medium Fermi energy $E_{\rm F,4}=k_{\rm F,4}^2/(2m_4)$. 
Therefore, $E_{\rm F,4}$ plays a crucial role in determining the critical temperature $T_{\rm c}$ of the Cooper triple condensation.
On the other hand, the estimation of $T_{\rm c}$ is non-trivial compared to the conventional BCS case because the mean-field Hamiltonian for the Cooper triple condensate cannot be diagonarized.
Nevertheless, by analogy with the BCS case, it may be reasonable to estimate $T_{\rm c}\sim \Lambda \exp\left(-\frac{1}{\rho(0)U_0}\right)$.
Also, the Gorkov-Melik-Barkhudanov-like correction would likewise suppress $T_{\rm c}$~\cite{GMB}.
By combining these expectations, we finally estimate $T_{\rm c}\sim 10^{-2}E_{\rm F,4}$, which will be able to be addressed in future experiments.\\

\noindent
{\it Conclusion and Outlook---}
In conclusion, we have elucidated how the Cooper triple condensation can occur in a three-component Fermi gas within the variational approach inspired by the BCS ground state for two-component fermions.
We have found that the Cooper triples can condense at zero center-of-mass momentum in the presence of a weak three-body attractive force among different components
as well as the Fermi surface common to the three components. The condensed state is predicted to have nontrivial degeneracy associated with the gauge and momentum orientations, which may induce interesting topological properties. Moreover, it is worth examining properties of excited Cooper triples, bound trimers in the strong coupling regime, lower dimensions, the influence of population imbalance,
and possible competition or coexistence between Cooper pairs and triples.
It is also interesting to consider superfluidity of the Coooper triple condensed state because the circulation of a singly quantized vortex would be different from the case of the Cooper pair condensed state by a factor of $2/3$.

This work was supported by Grants-in-Aid for Scientific Research from JSPS through Nos.\ 18H01211 and 18H05406.


\end{document}